\documentclass[final,3p,times,twocolumn,authoryear]{elsarticle}
\usepackage{lineno,hyperref}
\usepackage{graphicx}
\usepackage{adjustbox}
\usepackage{booktabs}
\usepackage{txfonts}
\usepackage{xcolor}
\usepackage{lipsum}
\usepackage{tabularx}
\usepackage{amsmath}
\usepackage{footnote}
\usepackage{soul}
\usepackage{multirow}
\usepackage{multicol}
\usepackage{mathpazo}

\modulolinenumbers[5]

\journal{Journal of \LaTeX\ Templates}

\biboptions{authoryear}

\begin{document}

\begin{frontmatter}

\title{The long-term X-ray flux distribution of Cygnus X-1 using  RXTE-ASM and MAXI observations}

\author[a]{Kabita Deka\corref{cor1}}
\ead{kabitadeka93@gmail.com}
\author[b]{Zahir Shah\corref{cor1}}
\ead{shahzahir4@gmail.com}
\author[b]{Ranjeev Misra}
\author[a]{Gazi Ameen Ahmed}

\address[a]{Department of Physics, Tezpur University, Napaam-784028, Assam, India}
\address[b]{Inter-University Center for Astronomy and Astrophysics, PB No.4, Ganeshkhind, Pune-411007, India}

\begin{abstract}
We studied the long term flux distribution of Cygnus X-1
  using RXTE-ASM lightcurves in two energy bands B (3-5 keV) \& C (5-12.1 keV) as well as MAXI lightcurves in energy bands B (4–10 keV) \& C (10–20 keV). The
  flux histograms were fitted using a two component model. For MAXI
  data, each of the
  components is better fitted by a log-normal distribution, rather
  than a Gaussian one. Their best fit centroids and fraction of
  time the source spends being in that component are consistent with
  those of the Hard and Soft spectral states
 Thus, the long term
  flux distribution of the states of Cygnus X-1 have a log-normal
  nature which is the same as that found earlier for
  much shorter  time-scales. For RXTE-ASM data, one component corresponding
  approximately to the Hard state is better represented by a log-normal
  but for the other one a  Gaussian is preferred and whose centroid 
  is not consistent with the Soft state. This discrepancy, could be due
  to the larger fraction of the Intermediate state ($\sim {11.25}$\%) in the
  RXTE-ASM data as compared to the MAXI one  ($\sim 4$\%). Fitting the
  flux distribution with three components did not provide an improvement for
  either RXTE-ASM or MAXI data, suggesting that the data corresponding to the Intermediate state
  may not represent a separate spectral state, but rather represent
  transitions between the two states.

\end{abstract}

\begin{keyword}
{methods: statistical -- accretion, accretion disks -- X-rays: binaries -- X-rays: individual: Cyg X-1 -- radiation mechanisms: thermal, non-thermal}
\end{keyword}

\end{frontmatter}

\nolinenumbers

\section{Introduction}\label{sec:intro}

Cygnus X-1 is one of the most powerful and persistent Galactic Black hole binary system, it was discovered during a rocket flight in 1964 \citep{Bowyer1965}. The system is a high mass X-ray binary with a black-hole of mass $21.2  \pm 2.2 \,M_\odot$ accreting matter from a super-giant O9.7 Iab star HDE\,226868 \citep{Miller2021, Gies1986, Herrero1995, Orosz2011}. The black hole and the star are in a quasi-circular orbit with an orbital period of  5.599829(16) days \citep{Brocksopp1999, Gies2003}.

 Cygnus X-1 is also one of the brightest X-ray sources in the sky. This source has been a prime target of most of the X-ray observations 
 from the time of its discovery. It shows strong variability at time scales from milliseconds to years at X-ray energies. Generally, the X-ray emission of Cygnus X-1 falls into one of the two distinct states, viz. ``low Hard" and ``high Soft" \citep{Liang1984}. This terminology is based on the long-term monitoring of Cygnus X-1 in the energy range 1--10 keV and it is used to describe the bi-modal behavior of the source. The X-ray emission during the hard state is  characterized by strong variability  ($>10\%$ rms) while in the  soft state, the variability  is  weak  ($ \leq 6\%$ rms \citep{Wilms2006}.
Prior to 2010, Cygnus X-1 spent most of the time in low Hard state \citep{Wilms2006,Grinberg2013}. However, the source behavior changed since 2010, it has spent significantly more time in Soft state \citep{Grinberg2013}. The Hard state is dominated by a non-thermal component, which is modeled by a power-law of index, $\Gamma$ $<$ 2.0 with an exponential cutoff at high energy $\sim$ 50--150 keV. While the Soft state is characterized by thermal dominated X-ray spectrum (i.e., a strong black body component with temperature,  $KT\sim 1$ keV) and a steep power law emission with index, $\Gamma>$ 2.5 \citep{Wilms2006, Grinberg2013}.
 In addition to Hard and Soft states, some times Cygnus X-1 is also found in the Intermediate state with X-ray spectral properties lying between the Hard  and Soft states \citep{1996ApJ...472L.107B}.  
The substantial difference in spectral and timing properties between the Hard and Soft states suggest for different accretion flow geometry and energetic. Most of the models proposed to explain the observed states contains two main components viz. geometrically thin optically thick disk, which produces the thermal radiation and an optically thin hot corona, which produces the non-thermal Comptonized radiation (e.g, \citealp{2002ApJ...578..357Z}).

Cygnus X-1 has been observed in different energy bands by various telescopes and dominant variability is reported in the X-ray light curves e.g. \citep{McHardy1988,Vaughan2003}. 
The variability is detected both in Hard and Soft states over time-scales ranging from seconds to years \citep{Frontera1975,Gleissner2004}. 
 Using the five years of MAXI data, \citet{2016PASJ...68S..17S} analyzed the long-term variations of Cyg X1 in the low/hard and high/soft state. The authors showed that the fractional RMS variation slightly decreases with energy in the low/hard state and increases towards higher energies in the high/soft state.
Also, the intensity modulation with the orbital period in  high/soft as well as in 
low/hard state is reported \citep{2017PASJ...69...52S}. 
Moreover \citet{2015A&A...576A.117G} showed  that the orbital variability is most prominent in the hard spectral state, 
they illustrated the variability in-terms of the probability distributions of absorption column density  values.
 \citet{Boroson2010} analyzed orbital variability due to absorption in five soft state periods by using RXTE-ASM light curves. Using the spectral state definition from \citet{Grinberg2013},  the authors showed the presence of orbital variability in soft state, while no such variability was found in the Intermediate state.

The  rapid X-ray variability of Cygnus X-1 shows characteristics of linear relationship between root mean square (r.m.s) variability and flux (linear r.m.s-flux relationship) i.e, absolute magnitude of r.m.s variability increases linearly with the mean flux level \citep{Uttley2001}. This linear rms-flux relation holds in all spectral states of Cygnus X-1 irrespective of their power spectral density (PSD) shape \citep{Gleissner2004}. Further, linear rms-flux relation has also been observed in other X-ray binary (XRB) systems  like neutron star XRB, Black-hole XRBs, ultra-luminous X-ray source (ULX) and Active Galactic Nuclei (AGN) like Seyferts,  blazars \citep{Uttley2001,Gaskell2004,Uttley2004,Uttley2005,Heil2010,Heil2011}. All these sources being powered by accretion suggest that this property is intrinsic to accreting systems.
\cite{Uttley2005} modeled the  observed linear r.m.s--flux relationship with a non-linear exponential model, which in-turn predicts the log-normal distribution of flux points in the light curve. 
Earlier, \cite{Negoro2002} had  shown that the distribution of peak X-ray intensities (shots) of Cygnus X-1 yields a log-normal distribution. 
Later regardless of shot modeling, \cite{Uttley2005} showed that  X-ray fluxes over shorter duration are better fitted with log-normal distribution.
Since log-normal distribution is an analogue of the normal distribution where the additive nature is replaced by the multiplicative one, observation of log-normal flux distribution/linear rms-flux relationship implies that the variability process is multiplicative rather than additive i.e., variation are coupled together on all time scales, thus placing strict constraint on the physical model responsible for  variability.
A propagating fluctuation model put forward by \cite{Lyubarskii1997} is one of the most  promising model which explains log-normal flux distribution/linear r.m.s-flux relationship. In this model, long-term variations from the outer regions of disk propagate inwards,  these modulations then couple with short-term variations at smaller radii of disk in a multiplicative way.

 Linear r.m.s-flux relationship in the X-ray light curves of Cygnus X-1 suggest a log-normal distribution of flux. \citet{2008MNRAS.389.1427P}  analyzed the Cygnus X-1 data corresponding to the hard spectral state, they showed that the X-ray flux on long timescales ($10^4$ s) is completely inconsistent with normal distribution and instead the flux distribution follows a log-normal  probability density functions (PDF). Earlier, \citet{Uttley2005} also showed that the hard state flux distribution on shorter time scales (1 s) follow a log-normal behavior. 
However, a detailed characterization of X-ray flux distribution of Cygnus X-1 on longer times scales including more than one spectral states has not been carried out.
In order to examine  this, we study the long-term X-ray flux distribution of Cygnus X-1 by using the RXTE-ASM 
and MAXI observations.  
 The aim of this work is to find-out whether different spectral states of Cygnus X-1 correspond to different probability distribution. Also, using the shape of probability distribution, we  examine the nature of Intermediate state i.e., whether the Intermediate state is distinct state or it is just the extreme version of the hard or soft flux states.

 In the next section, we describe the data selection for the RXTE-ASM
 and MAXI observation of Cygnus X-1 used in this work and the results
 of the flux distribution
 fitting are presented in Section \S 3.  In Section \S 4 we compare the
 distribution with the spectral state classification and the work
 is summarized and discussed in Section \S 5.

\section{RXTE-ASM and MAXI data selection:}
In order to understand the X-ray flux distribution characteristics of
Cygnus X-1, we used the publicly available long-term light curves of
Cygnus X-1 from RXTE-ASM and MAXI archive data base.

The Rossi X-ray Timing Explorer (RXTE) was launched  into low Earth orbit on December
30, 1995 from NASA's Kennedy Space Center.  
RXTE  carried an All-Sky Monitor (ASM;\cite{Levine1996} which monitored the 80\% of sky every 90 minutes in the energy range 1.5--12 keV.   
In this paper, we used  RXTE-ASM light curves of Cygnus X-1 publicly
available in the ASM/RXTE database
\footnote{http://xte.mit.edu/ASM$_-$lc.html. For RXTE ASM, the unit of Flux is  $count$ $sec^{-1}$. }, which is maintained by
the Massachusetts Institute of Technology (MIT).  These light curves
are obtained during the time period between January 5, 1996 (MJD
50087) to November 5,  2011 (MJD 55870), and  are available as 90 s
dwells and one-day averaged in three energy bands viz. A-Band (1.5-3.0
keV), B-Band (3.0-5.0 keV) and C-Band (5.0-12.1 keV)
 \citep{Levine1996}. Due to the
deterioration of ASM in the last two years of observation \citep{Grinberg2013, 2013MNRAS.428.3693V}, we carried the analysis by removing the ASM data following MJD 55200. Moreover,  the comprehensive study of RXTE-ASM dwell data showed that the orbital variability of absorption effects the hardness ratios viz B-band/A-band and C-band/B-band \citep{2000MNRAS.311..861B}. 
In order to check the effect of orbital variability on the daily binned light curves of RXTE-ASM, we used the definition of dips following \citet{2000MNRAS.311..861B}. We noted that the percentage of dips in the daily light curve are very less in C/B hardness ratio.  The decrease in the
 X-ray absorption with the increase in energy is also reported in \citet{1999ApJ...525..968W, 2006MNRAS.368.1025L}. Thus A-band being mostly affected by variable absorption, we considered only B-band and C-band light curves for our analysis.

To obtain reliable analysis, we imposed a criterion of signal to noise (SNR) for
each bin to be greater than 10 for both the B and C bands. We defined SNR
as the ratio of the counts to the error on counts, and eliminated any
time stamp if the SNR was less than 10 in either of the bands. Further, we
considered only the daily average data, since for the C-band dwell lightcurve, the error
fraction (i.e. the expected measurement variance divided by the
observed variance, $R =\overline{{\sigma^2_{err}}}/{\sigma^2}$ turned
out to be rather large $\sim 0.18$. For daily average light curve,  the error fraction for B and C bands are obtained as
$R \sim 0.002$ and $0.02$  respectively. The criteria SNR > 10,
removed $\sim 2.9$\% of the time bins and the final lightcurves used for analysis
consisted of 
4645 bins in each of the bands.

	In addition to RXTE-ASM light curves, we also examined the
publicly available daily light curves of Cygnus X-1 from  the Monitor of
All-sky X-ray Image (MAXI; \cite{Matsuoka2009}) on board the
International Space Station   which was developed by the Japan Aerospace
Exploration Agency (JAXA). It  is designed to monitor   X-ray
sources in the energy range 0.5--30 keV and it scans the full sky 
every 92 minutes.  MAXI uses two high sensitive X-ray cameras viz. the
Gas Slit Camera (GSC;\cite{Mihara2011}) and the Solid-state Slit
Camera (SSC; \cite{Tomida2011}).  The GSC surveys approximately 85\%
of the sky in one orbit in the energy range 2--30 keV
\cite{Sugizaki2011} while SSC operating in the night covers 30\% of
the sky.   Light curves from MAXI (GSC) are available in three energy
bands 2--4 keV (A-band), 4--10 keV (B-band) and 10--20 keV
(C-band)\citep{Matsuoka2009}. We retrieved the light curves in these
energy bands from the publicly  available MAXI RIKEN database \footnote{{ The data was retrieved during october 2018, it was processed with version 7L, where both 1650V and 1550V data are used ({\footnotesize {\tiny http://maxi.riken.jp/star$_-$data/J1958+352/J1958+352.html}}). 
 The unit of flux for MAXI is $photons$ $sec^{-1}$ $cm^{-2}$.}}. Again,
we consider B-band and C-band light curves for our analysis, since
the A-band may be affected by variable absorption. Like in RXTE light
curves, we removed the time bins simultaneously from B-band and C-band
for which error on counts is larger. For MAXI imposing a SNR > 10, removed
nearly 50\% of the time bins and hence we used a more lenient condition
that the SNR for each band is > 5, which removed $\sim 11$\% of the data.
The number of bin in final lightcurve was 1918 and the error fraction
was $R = 0.013$ and $0.062$ for the B and C bands respectively.

\section{Flux Distribution Characteristics} 

\begin{figure}
		\begin{center}
        
         \includegraphics[scale=0.5]{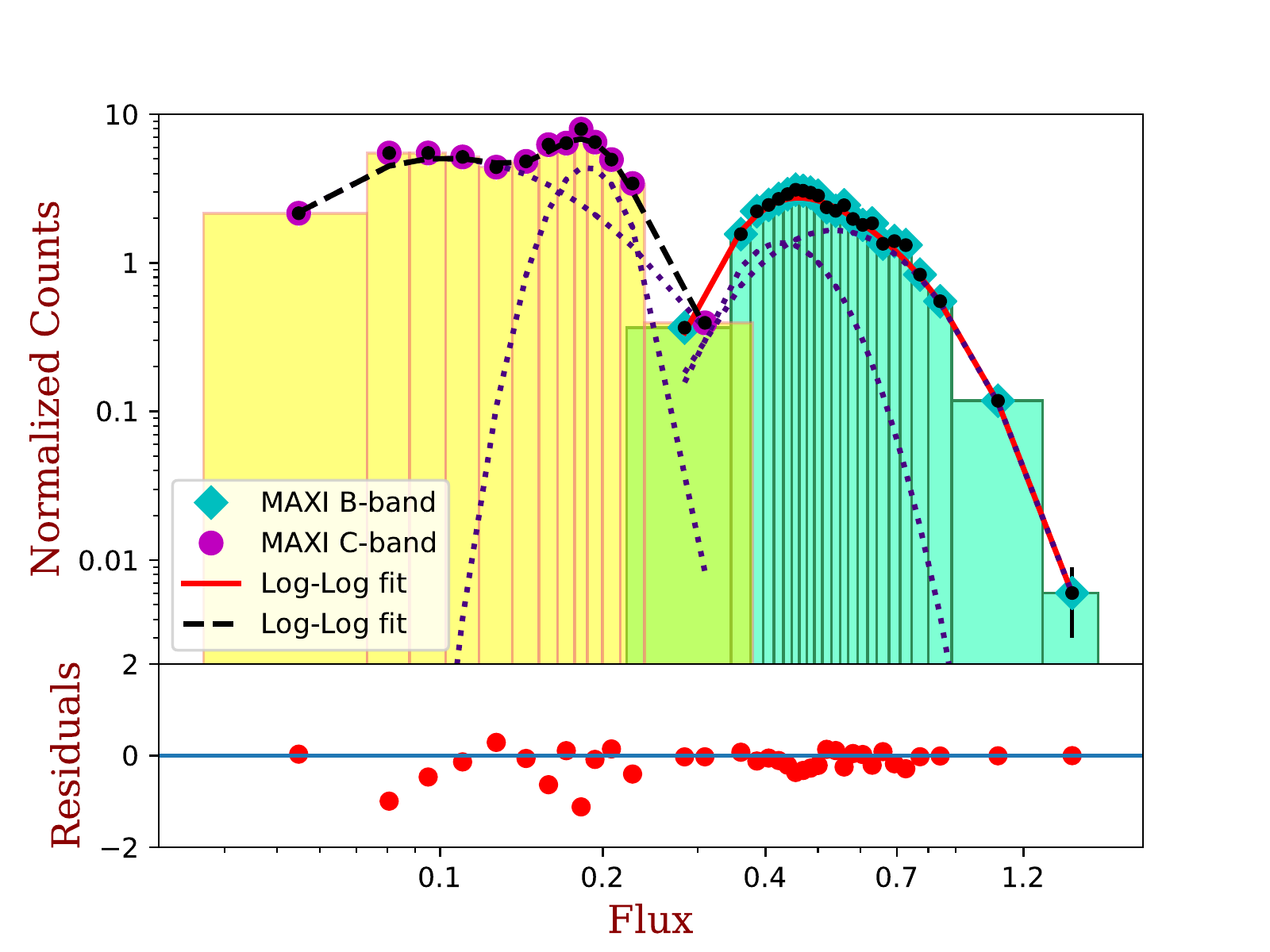}
        \caption{Top Panel : Histograms of the long-term X-ray flux distribution of Cygnus X-1 obtained using the MAXI B-band and C-band archive light curve. The green diamond points correspond to B-band light curves, while violet filled circles correspond to C-band light curve. Histogram with yellow and green colors corresponds to C-band and B-band respectively. The two curves (solid and dashed curves) shown are the best fitted double-log-normal PDF such that red solid curve corresponds to the B-band and black dashed curve corresponds C-band histogram. Dotted curves are the individual log-normal components. Bottom panel : Residuals}       
        \label{fig:maxi_hist}
		\end{center}        
        \end{figure}

\begin{figure}
		\begin{center}
       \includegraphics[scale=0.5]{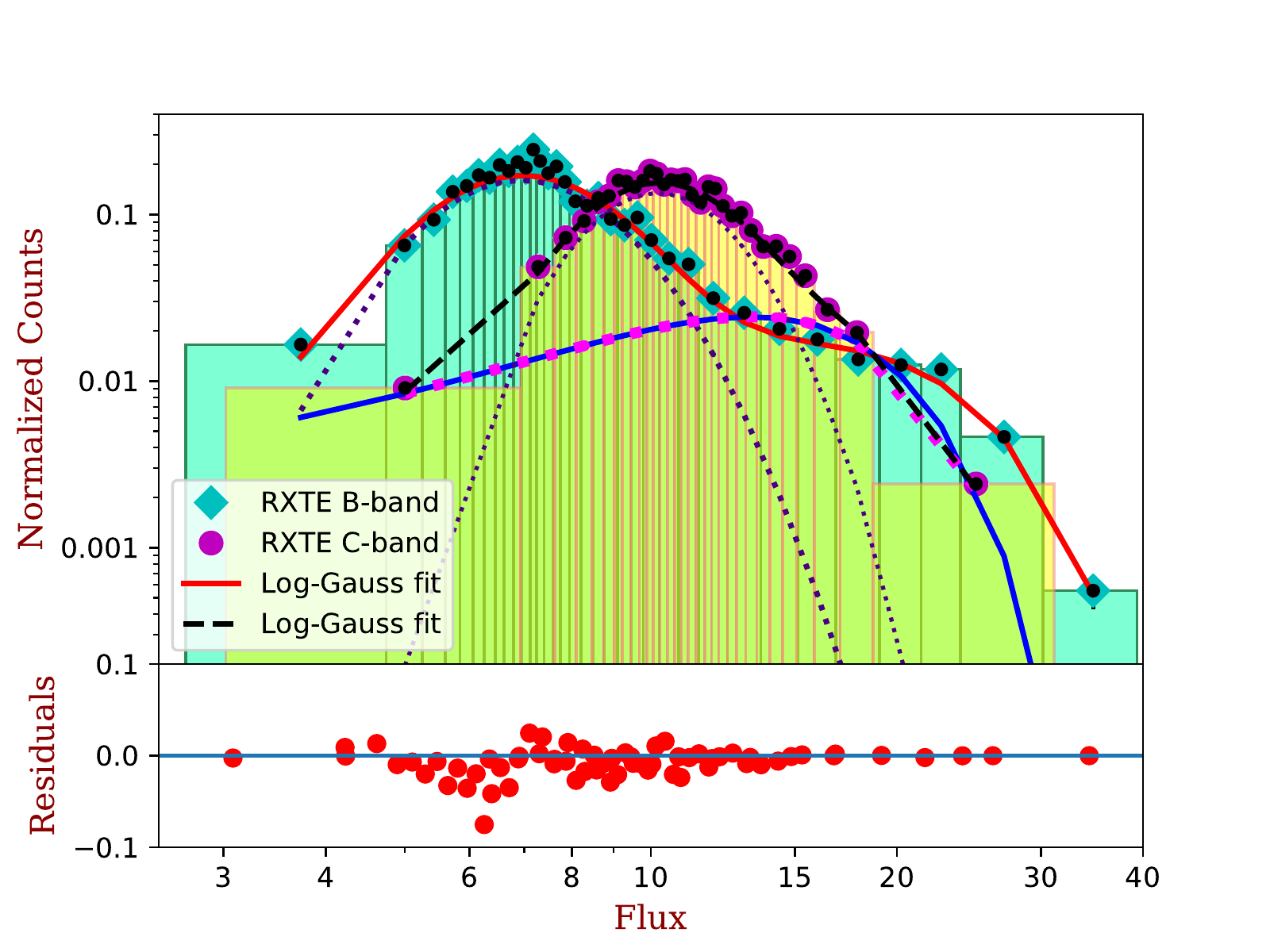}
        \caption{Top panel : Histograms of the long-term X-ray flux distribution of Cygnus X-1 obtained using the RXTE B-band and C-band archive light curve. The green diamond points correspond to B-band histogram, while violet filled circles correspond to C-band light curve. Histogram with yellow and green colors corresponds to C-band and B-band respectively. The two curves (solid and dashed curves) shown are the best fitted log-normal+Gaussian PDF such that red solid curve corresponds to the B-band and black dashed curve corresponds C-band histogram. Dotted curves are the individual log-normal components. Solid(Blue) and dotted(fuchsia) curves are the individual Gaussian components for B and C bands respectively. Bottom panel : Residuals }
        \label{fig:rxte_hist}
		\end{center}        
        \end{figure}

 Since Cygnus X-1 is known to have at least two distinct spectral states,
 its flux distribution would require more than one component.
 For completeness, we verified using the Anderson-Darling and skewness tests that neither a single Gaussian or a log-normal distribution is consistent with
 the flux distribution.
 In order to characterize the probability distribution, we created normalized histograms of  RXTE-ASM/MAXI B and C band light curves. The  histogram binning was chosen such that each bin in the histogram contained equal number of flux points.
 The resultant histogram obtained for the MAXI (B-band and C-band) and RXTE-ASM (B-band and C-band) light curves are shown in Figure \ref{fig:maxi_hist} and \ref{fig:rxte_hist}.
 Now in order to reproduce the shape of these histograms, we define two PDF's as

\begin{eqnarray}\label{eq:ln}
L(x)=\frac{1}{{\sqrt{2\pi}\sigma_l x}}\exp\left({\frac{-[log(x)-{\mu_l}]^2}{2{\sigma_l}^2}}\right)
\end{eqnarray}
and

\begin{eqnarray}\label{eq:gaus}
G(x)=\frac{1}{{\sigma_g}{\sqrt{2\pi}}}\exp\left(\frac{-[x-{\mu_g}]^2}{2{\sigma_g}^2}\right)
\end{eqnarray}

 Equation \ref{eq:ln} results in a log-normal fit with  $\mu_l$ and $\sigma_l$ as the centroid and width of corresponding logarithm flux distribution while Equation \ref{eq:gaus} results in a Gaussian/normal fit with $\mu_g$ and $\sigma_g$ as the centroid and  width of corresponding normal distribution.
 For the double PDF fit, we chose different combinations like log-normal-Gaussian (Equation \ref{eq:ln} \& \ref{eq:gaus}), log-normal-log-normal (Equation \ref{eq:ln} and \ref{eq:ln}).  For example, the double-log-normal PDF  is given by

\begin{eqnarray}\label{eq:d_ln}
D(x)  =L_1(x)+L_2(x) \nonumber \\ 
	=\frac{f_h}{\sqrt{2\pi}\sigma_{l1} x}\exp\left({\frac{-[\log(x)-{\mu_{l1}}]^2}{2{\sigma_{l1}}^2}}\right) \nonumber \\
	 +\frac{1-f_h}{\sqrt{2\pi}\sigma_{l2} x}\exp\left({\frac{-[\log(x)-{\mu_{l2}}]^2}{2{\sigma_{l2}}^2}}\right)\nonumber
\end{eqnarray}

 where $\mu_{l1}$ and $\mu_{l2}$ are the centroids of the logarithm flux distributions with widths $\sigma_{l1}$ and $\sigma_{l2}$, respectively, $f_h$ is fraction of time the source is in Hard state, we nominally associate the lower flux component of the B-band with the Hard state.
Since, Cygnus X-1 is observed simultaneously in B-band and C-band, and the fraction of time it spends in Hard/Soft state is same for both bands. We therefore carried simultaneous fitting of B  and C band histograms with double PDF by keeping `$f_h$' same.  This will ensure similar contribution of the respective components from B and C bands. Note that for RXTE B and C bands, and MAXI B-band, the lower flux levels correspond to the Hard state. On the other hand, the Hard state of MAXI C-band corresponds to the higher flux levels.
The best fit parameter values acquired by using different double PDF for
MAXI data are summarized in Table \ref{table:maxi_dpdf}.
The reduced-$\chi^2$ ($\chi^2_{red}$) obtained  suggest that the MAXI B and C band histograms are better fitted with double log-normal PDF as shown in figure \ref{fig:maxi_hist}. For RXTE-ASM the histogram fitting resulted in a large
$\chi^2_{red}$, In case of RXTE data, the double PDF fit to the B and C band results in large reduced-$\chi^2$ ($\chi^2_{red}$ >2).
The fit was not improved even with the tri-model PDF fit. Therefore we introduced a 3\% systematics to
obtain a better reduced-$\chi^2$ ($\chi^2_{red}$). In previous study, the 3\% systematic error is introduced in the light curve
of Cygnus X-1 based on Crab pulsar measurements\citep{Boroson2010},
and the corresponding best fit values are  listed in Table \ref{table:rxte_dpdf}. For RXTE-ASM data, a combination of log-normal and Gaussian seems to fit
the data better than a double log-normal one and the fitted histograms are
shown in Figure \ref{fig:rxte_hist}.



\begin{table*}
\centering
\caption{Best fit parameter values of  double PDF's fitted to the flux histograms of MAXI B \& C band.
 Col:- 1: PDF , 2-9: Best fit values of $\mu_{1}$, $ \sigma_{1}$, $\mu _{2} $, $ \sigma_{2}$, $ \mu_{11}$, $ \sigma_{11}$, $\mu _{21} $ and $ \sigma_{21}$, 10: fraction parameter and  11: Reduced $\chi^2$ (degrees of freedom )}
\vspace{0.2cm}
\begin{adjustbox}{width=1.0\textwidth,center=\textwidth}
\begin{tabular}{lcccccccccr}
\cline{1-11}
MAXI & \multicolumn{4}{c}{B band} & \multicolumn{4}{c}{C band}  \\  
\cline{1-11}
 PDF &  $\mu_{1}(\times10^{-1})$ & $\sigma_{1}(\times10^{-1})$ & $\mu _{2}(\times10^{-1}) $ & $ \sigma_{2}(\times10^{-1})$ &  $\mu_{11}(\times10^{-1})$ & $\sigma_{11}(\times10^{-1})$ & $\mu _{22}(\times10^{-1}) $ & $ \sigma_{22}(\times10^{-1})$ & $f_h$ & $\chi_{\rm red}^2$(dof) \\
 \cline{1-11}
Log \& Log & 4.4$\pm$0.13 & 1.9$\pm$0.12 & 5.8$\pm$0.05 & 3.0$\pm$0.12 & 1.2$\pm$ 0.01 & 4.8$\pm$0.12 & 1.9$\pm$0.01 & 1.4$\pm$0.13 & 0.29$\pm$0.02 & 1.06(26) \\

 Log \& Gauss & 4.6$\pm$0.04 & 1.6$\pm$0.11 & 6.0$\pm$ 0.08 & 2.0$\pm$0.06 &  1.2$\pm$0.01 & 4.9$\pm$0.14 & 1.8$\pm$ 0.02 & 0.2$\pm$0.02 & 0.34 $\pm$0.02 & 1.96(26) \\

 Gauss \& Gauss & 4.5$\pm$0.05 & 0.6$\pm$0.06 & 6.0$\pm$0.08 & 1.9$\pm$0.05 &  0.8$\pm$0.02 & 0.2$\pm$0.02 & 1.8$\pm$ 0.03 & 0.5$\pm$0.01 & 0.27 $\pm$0.02 & 3.33(26) \\
	\cline{1-11}
\end{tabular}
\end{adjustbox}
\label{table:maxi_dpdf}
\end{table*}

\begin{table*}

\centering
\caption{Best fit parameter values of double PDF's fitted to the flux histograms of RXTE B \& C band.  Col:- 1: PDF , 2-9: Best fit values of $\mu_{1}$, $ \sigma_{1}$, $\mu _{2} $, $ \sigma_{2}$, $ \mu_{11}$, $ \sigma_{11}$, $\mu _{21} $ and $ \sigma_{21}$, 10: fraction parameter and  11: Reduced $\chi^2$ (degrees of freedom )}\label{tlc}
\vspace{0.2cm}
\centering
\begin{adjustbox}{width=1.0\textwidth,center=\textwidth}
{\begin{tabular}{lcccccccccr}
\cline{1-11}
RXTE & \multicolumn{4}{c}{B band} & \multicolumn{4}{c}{C band}  \\ 	
\cline{1-11}
PDF &  $\mu_{1}$ & $\sigma_{1}(\times10^{-1})$ & $\mu _{2} $ & $ \sigma_{2}$ &  $\mu_{11}$ & $\sigma_{11}(\times10^{-1})$ & $\mu _{22} $ & $ \sigma_{22}$ & $f_h$ & $\chi_{\rm red}^2$(dof) \\
 \cline{1-11}
Log \& Gauss & 6.42 $\pm$0.05 & 2.6$\pm$0.09 & 13.32$\pm$0.90 & 8.00$\pm$0.43 & 9.48$\pm$0.08 &  2.1$\pm$0.08 & 12.17$\pm$0.31 & 5.45$\pm$ 0.26 & 0.66 $\pm$0.02 & 1.20(55) \\

Log \& Log & 6.48$\pm$0.06 & 2.9$\pm$0.09 & 16.94$\pm$0.67 & 0.31$\pm$0.02 & 9.87$\pm$ 0.08 & 2.4$\pm$0.08 & 11.02$\pm$0.04 & 0.54$\pm$0.04 & 0.78$\pm$0.01 & 1.55(55) \\

Gauss \& Gauss & 6.42 $\pm$0.07 & 1.50$\pm$0.07 & 13.08$\pm$1.22 & 7.70$\pm$0.63 & 9.37$\pm$0.12 &  1.91$\pm$0.11 & 12.96$\pm$0.56 & 4.81$\pm$ 0.26 & 0.62 $\pm$0.03 & 2.70(55) \\
\cline{1-11}
\end{tabular}}
\end{adjustbox}
\label{table:rxte_dpdf}
\end{table*}


\section{Comparison with Spectral States}
In addition to Hard and Soft spectral states, Cygnus X-1 also exhibits an Intermediate spectral state characterized by a moderately strong thermal component and relatively soft Hard spectrum \citep{1996ApJ...472L.107B}. The Intermediate state mostly appears when the source is about to make a transition from Hard  to Soft state, but fails to do and instead stays in a distinct state, where the X-ray spectral properties lies in between Hard state and Soft state \citep{1996ApJ...472L.107B,2003A&A...407.1039P}. \citet{Grinberg2013} have
provided an easy-to-use prescription in determining the spectral states of Cygnus X-1 using data from RXTE-ASM and MAXI.
For convenience, we have repeated the state definitions of Hard, Soft and Intermediate states of Cygnus X-1 in Table \ref{Table:state_def}. Using these definitions, we separated the B-band and C-band data of RXTE-ASM and MAXI into Hard, Soft and Intermediate states. The Table also shows the number of observations per state and the corresponding percentage.

 For MAXI data, there is a interesting correspondence between the Hard and Soft states and the flux distribution components  obtained in the simultaneous fitting of B
and C band histograms with double PDF. 
 The best fit centroid values of the two components
 matches closely with the average of the fluxes\footnote{We computed the average of the logarithm
of the flux for a state when we are comparing it with the log-normal
distribution and the average of the flux when we compare with a Gaussian one.} of the Hard and Soft states (see table \ref{table:mean_val}). We note that for MAXI observations, the fraction of time the source is in the Intermediate State is $\sim 4$\% and hence including them as being part of either the Hard or Soft state, does not change the flux averages or their fraction of time spent in them. However, the fraction of time spend in the Hard state based on spectral classification ($\sim 39.8$\%) deviates from that obtained by flux distribution fitting $f_{h} = 29\pm2$\% (see table \ref{table:maxi_dpdf}). It should be noted here that in case of MAXI, the criterion $SNR > 5$ for each time bin in both the B and C bands removes significant number of timebins (11\% of the total data). We found these timebins gets preferentially removed from the hard state. 
This means that the obtained $f_{h} = 29\pm2\%$ value would be significantly lower than the actual time spend by the source in the hard state. If we consider that the 10\% of removed data belongs to the hard state, then $f_h$ would turn out to be $\sim 39\%$, which is much close to the value obtained from the spectral classification ($\sim 39.8\%$).

For RXTE  data, 
the best fit centroid value and the average flux value for the hard state matches well with that
 found by the flux distribution fitting as shown in Table \ref{table:mean_val}., but a deviation can be seen  for the second component i.e. with the Soft state.
 Here, the Intermediate state is represented by a significant fraction of the observations $\sim 11$\% comparable to the Soft state. 
As shown in Table \ref{table:mean_val},
 combining the Intermediate and the Soft state observations does not alleviate the discrepancy between the spectral and
 flux distribution classification. Moreover, we found that the fraction of time spend in the Hard state based on spectral classification ($\sim 77.45$\%)  also deviates from that obtained by flux distribution fitting $f_{h} = 66\pm2$\% (see table \ref{table:rxte_dpdf}). However, in case of RXTE, only small fraction of timebins  (2.9\%) gets removed by SNR criteria, and therefore the discrepancy can not accounted for the timebins removed by SNR criteria.

 We consider the possibility that the flux distribution should be represented by three components corresponding to the
 Hard, Soft and Intermediate spectral states. Thus, we attempted to fit triple combinations of log-normal and Gaussian
 to the histograms. Since the number of free parameters are large, we constrain the centroids of two of the components
 to match with the average fluxes of the Soft and Intermdiatory states. As shown in Table  \ref{Table:triple_fit}), for
 various combinations, the $\chi^2_{red}$ is not smaller than what was obtained for the two component fit and hence we infer
 that there is no evidence for the flux distribution to be represented by three components each being either a log-normal
 or a Gaussian.

\begin{figure}
		\includegraphics[height=9cm, width=9cm]{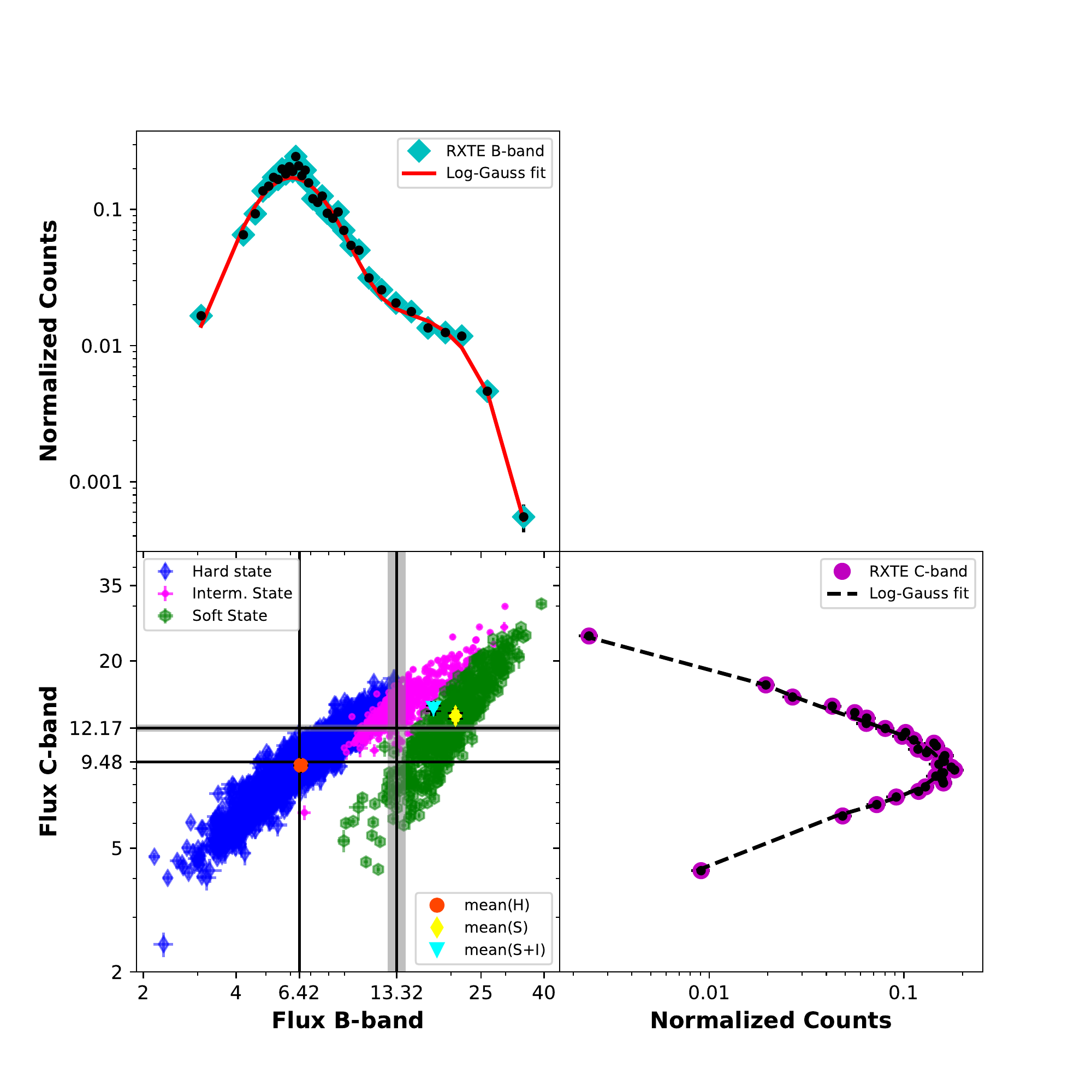}
				\caption{Multiplot of Cygnus X-1 obtained using the RXTE-ASM observations. The top and bottom-right panels are the histograms of long-term X-ray flux distributions of B and C band archive light curves. The curves (red and black-dashed curves)  are the best fitted double PDF (log-normal+Gaussian). Bottom-left panel is the plot between the  B-band and C-band flux's with the Hard, Inter-mediate and Soft states highlighted by colors blue, orange and green respectively. The vertical and horizontal gray bands indicate the centroids of log-normal+Gaussian PDF fit to the histograms of B and C bands with widths representing 1-$\sigma$ error range. Large size points with colors red, cyan and yellow mark the mean flux points of Hard, Soft+Intermediate and Soft states respectively.}
\label{fig:state_def_rxte}
\end{figure}

\begin{figure}
		\includegraphics[height=9cm, width=9cm]{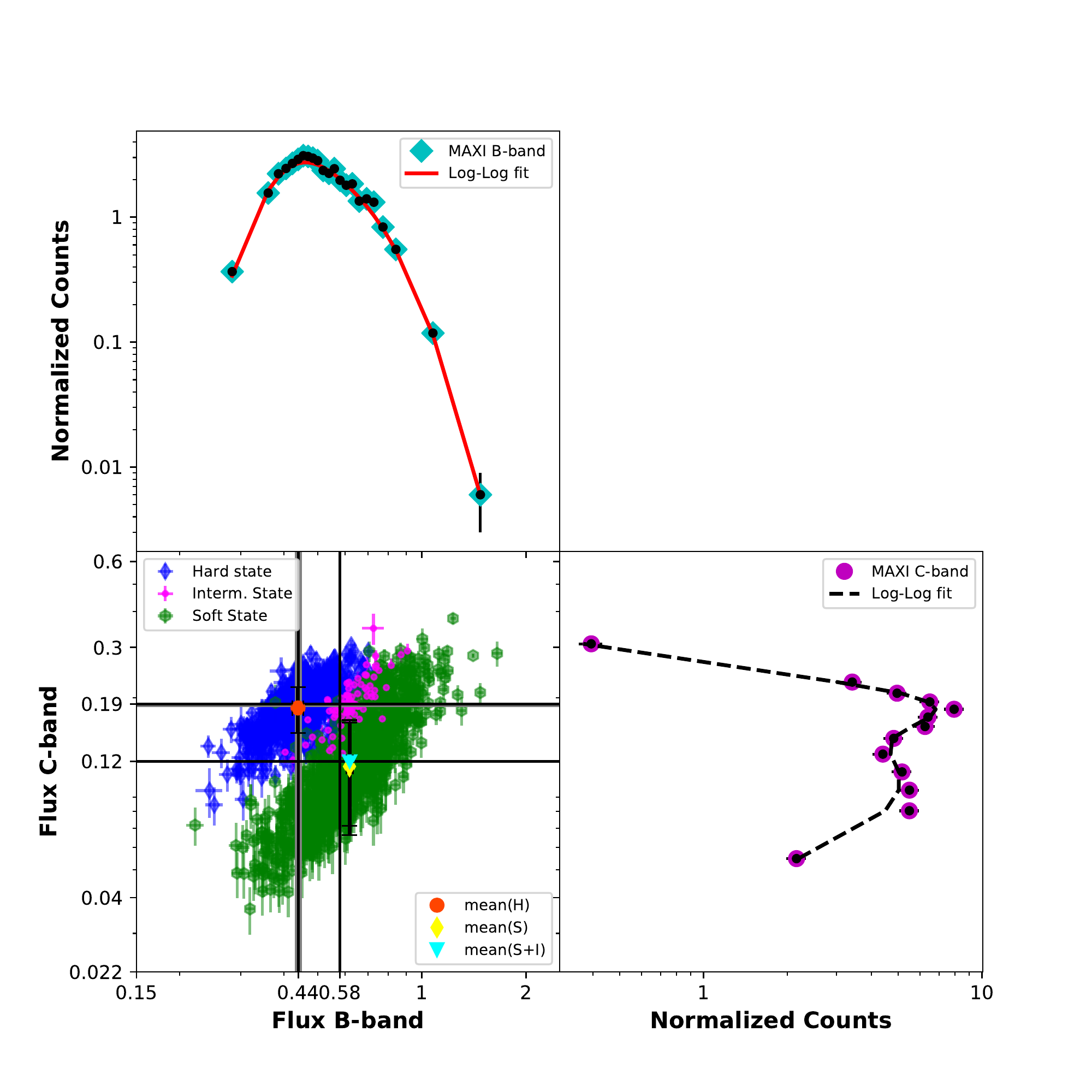}
				\caption{Multiplot of Cygnus X-1 obtained using the MAXI observations. The top and bottom-right panels are the histograms of long-term X-ray flux distributions of B and C band archive light curves. The curves (red and black-dashed curves) are the best fitted double log-normal PDF'.  Bottom-left panel is the plot between the  B-band and C-band flux's  with the Hard, Inter-mediate and Soft states highlighted by colors blue, orange and green respectively. The vertical and horizontal gray bands indicate the centroids of double-log-normal PDF fit  to the histograms of B and C bands respectively with widths representing 1-$\sigma$ error range. Large size points with colors red, cyan and yellow represent the mean flux points of Hard,  Soft+Intermediate and Soft states respectively.}
\label{fig:state_def_maxi}
\end{figure}

The results of this work can be graphically summarized by multiplots
as shown in Figure  \ref{fig:state_def_rxte} for RXTE/ASM
and Figure \ref{fig:state_def_maxi} for MAXI. The top and right panels
are the flux histogram with the best fit two component models.
In the central panel, the C-band and B-Band fluxes are plotted against
each other and the points corresponding to the Hard, Soft and
Intermediate States (as defined by \citet{Grinberg2013}) are shown using
different point types (and color). The vertical and horizontal
solid bands represent the centroid values of the components used
to fit the flux histograms with the widths equal to their errors.
Large size points mark the average
flux values of the Soft, Hard and Soft+Intermediate
states. For both energy bands and for both the instruments, the
average of the Hard state coincides with the best fit centroid
values. The average of the Soft+Intermediate states are closer to
the corresponding centroid values as compared to the Soft state alone.
\begin{table*} 

\caption{The state definitions of RXTE-ASM data and MAXI described by \citealp{Grinberg2013}. Col:- 1: States of Cygnus X-1, 2: Criteria for identifying spectral states of Cygnus X-1 using RXTE-ASM observation, 3: Criteria for identifying states of Cygnus X-1 for MAXI observations,  4 \& 5: Number of data points, N with their percentage that fall into each of the states using RXTE-ASM and MAXI observation. In Col 2,  c is count rate (counts $\rm s^{-1}$) in energy band $\sim$1.5-12 keV,  h is hardness ratio defined as ratio of count rates in RXTE C-band ($5-12$ keV) and A-band ($1.5-3$ keV) and $h_{0}$=0.28.  In Col 3,  $c_{M}$ is  count rate in energy range $\sim 2-4$ keV,  $h_{M}$ is MAXI hardness defined as $4-10$ keV / $2-4$ keV (B-band/A-band)}
\vspace{0.2cm}

\begin{adjustbox}{width=1.0\textwidth,center=\textwidth}

\begin{tabular}{lcccc} 
\cline{1-5}
State & RXTE-ASM & MAXI & N(Percentage)  &  N(Percentage)\\
&	&	&	using RXTE-ASM	&  using MAXI\\
\cline{1-5}
Hard state & c $\leq$ 20 $\vee$ c $\leq$ 55 $\times(h-h_{0})$ & $c_{M}$ $\leq$ 1.4$\times(h_{M})$ &  3598(77.45\%) & 763(39.78\%)\\ 
Interm. state & c $>$ 20 $\wedge$ 55 $\times(h-h_{0})$  $<$ c $\leq$ 350 $\times(h-h_{0})$ & 1.4$\times(h_{M})$ $<$ $c_{M}$ $\leq$ 8/3$\times(h_{M})$ & 523(11.25\%) & 81(4.22\%)\\
Soft state & c $>$ 20 $\wedge$ c $>$ 350 $\times(h-h_{0})$ & 8/3$\times(h_{M})$ $<$ c & 524(11.28\%) & 1074(55.99\%)\\ 
\cline{1-5}
 \end{tabular}
\end{adjustbox}
\label{Table:state_def}

\end{table*}

\begin{table*}
\caption{Centroid values obtained from Double PDF's fitting and Mean values of different spectral states of B and C bands of RXTE and MAXI.}
\vspace{0.2cm}
\begin{adjustbox}{width=1.0\textwidth,center=\textwidth}
\begin{tabular}{lccccccc}
\cline{1-8}
& \multicolumn{2}{c}{Centroid Values of PDF} & \multicolumn{5}{c}{Mean of Spectral State} \\
\cline{1-8}
Energy band & $\mu_H$ & $\mu_S$ & mean(H) & mean(S) & mean(I) & mean(H+I) & mean(S+I)\\
\cline{1-8}
RXTE B-band & 6.42$\pm$0.05 & 13.32$\pm$0.90 & 6.49$\pm$0.01 & 20.08$\pm$0.01 & 14.01$\pm$0.01 & 7.15$\pm$0.01 & 16.78$\pm$0.01 \\
RXTE C-band & 9.48$\pm$0.08 & 12.17$\pm$0.31 & 9.23$\pm$0.01 & 12.66$\pm$0.01 & 14.62$\pm$0.01 &	9.78$\pm$0.01 & 13.60$\pm$0.01 \\
MAXI B-band & 0.44$\pm$0.02 & 0.58$\pm$0.005 & 0.43$\pm$0.01 & 0.61$\pm$0.01 & 0.62$\pm$0.01 & 0.45$\pm$0.01 & 0.61$\pm$0.01 \\
MAXI C-band  & 0.19$\pm$0.01 & 0.12$\pm$0.001 & 0.18$\pm$0.03 & 0.11$\pm$0.04 & 0.19$\pm$0.13 & 0.18$\pm$0.03 & 0.11$\pm$0.04 \\
\cline{1-8}
\end{tabular}
\end{adjustbox}
\label{table:mean_val}
\end{table*}

\begin{table*}
\caption{Best fit parameter values of the triple PDF's fitted to the flux histograms of RXTE and MAXI B \& C bands.
Col:-1: Combination of functions , 2-9: Best fit values of $\mu_{1}$, $ \sigma_{1}$, $\sigma _{2} $, $ \sigma_{3}$ , $\mu_{11}$ , $\sigma_{11}$ , $\sigma _{22} $ \& $ \sigma_{33}$, 10-11: Normalization fractions $f_{n1}$ \& $f_{n2}$ and  12: $\chi_{\rm red}^2$(Degrees of freedom )}
\vspace{0.2cm}
\centering
\begin{adjustbox}{width=1.0\textwidth,center=\textwidth}
\resizebox{\textwidth}{!}
{\begin{tabular}{lccccccccccr}
 \cline{1-12}
RXTE & \multicolumn{4}{c}{B band} & \multicolumn{4}{c}{C band} &  \multicolumn{2}{c}{Norm. Fraction}  &  \multicolumn{1}{c}{Red. $\chi^2$}\\
 \cline{1-12}
 PDF &  $\mu_{1}$ & $\sigma_{1}(\times10^{-1})$ & $\sigma _{2} $ & $ \sigma_{3}$ &  $\mu_{11}$ & $\sigma_{11}(\times10^{-1})$ & $\sigma _{22} $ & $ \sigma_{33}(\times10^{-1})$ & $f_{n1}$ & $f_{n2}$ & $\chi_{\rm red}^2$(dof) \\
  \cline{1-12}

Log \& Gauss \& Gauss &  6.48$\pm$0.06 & 2.9$\pm$0.08 & 4.22$\pm$1.22  & 6.17$\pm$0.45 &  9.39$\pm$0.09 & 0.21$\pm$0.09 & 2.27$\pm$ 0.40  & 6.87 $\pm$0.89 & 0.77 $\pm$0.01 & 0.1$\pm$0.03  &1.69(55) \\

Log \& log \& log  &  6.48$\pm$0.06 & 2.9$\pm$0.07 & 0.30$\pm$0.05  & 0.25$\pm$0.01 &  9.48$\pm$0.09 & 2.2$\pm$0.08 & 0.17$\pm$ 0.03 & 0.92 $\pm$0.24 & 0.77 $\pm$0.01 & 0.1$\pm$0.02  & 1.34(55) \\
 \cline{1-12}
\\
 MAXI \\
 \cline{1-12}

Log \& log \& log  &  0.44$\pm$0.01 & 0.21$\pm$0.01& 0.23$\pm$0.11  & 0.27$\pm$0.01 &  1.0$\pm$0.05 & 0.35$\pm$0.03 & 0.12$\pm$0.05 & 0.20 $\pm$0.01 & 0.45 $\pm$0.05 & 0.09$\pm$0.05  & 1.17(25) \\

Log \& gauss \& gauss  &  0.45$\pm$0.01 & 0.17$\pm$0.02 & 0.13$\pm$0.01  & 0.24$\pm$0.02 &  1.0$\pm$0.04 & 0.36$\pm$0.03 & 0.08$\pm$0.01 & 0.02 $\pm$0.03 & 0.34 $\pm$0.06 & 0.35$\pm$0.11  & 1.37(25) \\
 \cline{1-12}
 
\end{tabular}}
\end{adjustbox}
\label{Table:triple_fit}
\end{table*}

\section{Summary and Discussion}

We have analyzed the long term flux distribution of Cygnus X-1 using
RXTE/ASM and MAXI data in two energy bands. We fit the
distributions using two components which we consider to be
either a log-normal or a Gaussian one.
For MAXI data, both the components are better represented by
a log-normal distribution, rather than a Gaussian one, and they
can be clearly identified with the Hard and Soft spectral states.
 The centroid values of the two components and the fraction of time they occur are consistent
with the independent spectral-timing classification undertaken
by  \citet{Grinberg2013}. Thus, this implies that the
long term flux distribution of Cygnus X-1  is
the same as for its short term one. From time-scales of seconds
to week, the variability of the source  seems
to be determined by an multiplicative process leading to log-normal
distributions. It is attractive to associate a single process such
as a stochastic propagation model where perturbations in different
time-scales originating at different radii of the disk have a
multiplicative effect on the inner accretion rate leading to X-ray
variability  \citep{Lyubarskii1997}. However, there are other possibilities such
as Gaussian variability of spectral index \citep{2018MNRAS.480L.116S} and emission from large number of randomly oriented mini-jets \citep{2012A&A...548A.123B}
which can give rise to a log-normal flux distribution and it is
not certain whether these different mechanisms are active on different
time-scales.

For RXTE/ASM data, one of the components is better described as a
log-normal one and its centroid value and fraction of occurrence
time is close to that obtained from spectral classification of
 \citet{Grinberg2013} for the Hard state. For MAXI
  data, each of the
  components is better fitted by a log-normal distribution. Their best fit centroids and fraction of
  time the source spends being in that component are consistent with
  those of the Hard and Soft spectral states. Thus, the MAXI result,
that the Hard state's long term distribution is a log-normal one, is
corroborated by the RXTE data. However, for the second distribution
the results are in contradiction since the best fit obtained is
a Gaussian function instead of a log-normal one. Further,
the centroid of the component does not match well with the average
flux of the Soft state nor of the Soft and Intermediate states together.


If we consider the Intermediate state as a distinct state as defined by \citet{Grinberg2013} than one obvious reason for the discrepancy between the MAXI and the RXTE/ASM
results might be  the presence of observations classified as Intermediate state.
For MAXI data such observations only account for $\sim 4$\% of the
total data while for RXTE/ASM it is significantly larger at $\sim 11.25\%$
and comparable to the Soft State fraction of $\sim 11.28$\%.
To check whether observations showing Intermediate spectral and temporal
behavior belong to a separate state or not, we attempted
to fit the long term flux histograms with three components. Since the
number of parameters for a three component fit (thirteen) was too many given
the data set, we took guidance from the spectral classification by  \citet{Grinberg2013}
to fix the centroid fluxes for the soft and Intermediate components to
the mean values of these classes. The reduced $\chi^2$ obtained were larger than
those obtained for a two component fit for both ASM and MAXI data. Thus,
we were not able to find any indications for the presence of three flux
distribution components. It should be noted that if  the
Intermediate state observations correspond to times when the source
is making a transition between the Soft and Hard states (i.e. it is
not a distinct state by itself) then their flux distribution may not
correspond to a Gaussian (or log-normal) and instead could be some
arbitrary function. In that case, flux distribution fitting may
not reveal their nature since such complex triple component fits may always
be degenerate.

The two spectral components that dominate
the Soft state spectra (i.e. a thermal disk emission and a power-law
Comptonization one) may have different distributions. We tried to check this
by fitting the soft state flux distributions by a Gaussian for  the MAXI
B band and by a log-normal for the MAXI C band and vice-versa
However, the reduced
$\chi^2$ for a double log-normal description was found to be smaller
that such combinations. Since the C-band may be dominated by the
power-law emission and the B-Band by the thermal one, this implies that both the thermal disk emission
and the power-law one follow log-normal distributions.

Analysis of a large number of pointed
observations by a more sensitive instrument like the RXTE/PCA will provide
significantly better information than the sky monitor data used here.
Moreover, such an analysis can be undertaken for finer energy bins
to quantify differences, if any, between the flux distributions of different
spectral components. However, for such an analysis, the serious
bias introduced by uneven sampling
and by observations preferentially undertaken when the source was in an unusual
state has to be addressed.

The flux distribution analysis shows that fraction of time Cygnus X-1 was in a Hard state during the RXTE era, $\sim 0.7$,  was significantly
higher than during the MAXI observations, $\sim 0.3$. This implies behavioral changes on time-scales of decades. 
 The change in spectral state of Cygnus X-1 in recent years has been reported in earlier studies as well \citep{Grinberg2013,2016PASJ...68S..17S}. 
It is interesting
to note that there are black hole systems (such as LMC X-1) which are always observed to be in the Soft state and it maybe possible that
after
many decades,  a system like Cygnus X-1 may evolve, such that it also spends most of its time in one particular state. 
We emphasis that classification based on spectral and timing properties of
a source provide valuable insight into the nature of the system and in particular are extremely useful in making predictions for the source behavior. On the
other hand, flux distribution analysis reveal the number of distinct dynamic
configurations (i.e. broadly corresponding to different flux levels)
a system can be in. If the two analysis are consistent as seen for the
MAXI data, then
it reinforces the distinct spectral/timing properties of a system for a
given configuration. These configurations may correspond to hydro-dynamical
steady state solutions of accreting systems which co-exist for a given
determining parameter such as the accretion rate or the system may
make transition from one solution to another when the parameter changes \citep{1976ApJ...204..187S, 1994ApJ...428L..13N}.

Unlike Cygnus X-1 which is persistent, many black hole systems exhibit
outburst which last for several months and which occur on once in several
years. They show a wide range of flux levels with different spectral and
timing properties. A flux distribution analysis will require observations
of a significant number of outbursts, otherwise the analysis will be biased
by a few dominant ones. On the other hand, flux distribution studies can be
undertaken on persistent sources (including  Neutron star systems) using monitoring
instruments with repeated and unbiased pointed observations.

\section{ACKNOWLEDEMENTS}
K. Deka would like to thank University Grants Commission of India and Tezpur University for all the support. K. Deka acknowledges the support of IUCAA under the visitor’s program, most part of this research work is carried under this program. G.A. Ahmed would like to thank IUCAA for associateship. This research work has made use of MAXI data provided by RIKEN, JAXA, and the MAXI team, and RXTE-ASM light curves provided by the ASM/RXTE team.


\section*{}

\bibliography{kabita.bib}

\end{document}